\documentclass[11pt,a4paper]{article}
\pdfoutput=1
\usepackage{jheppub}

\newcommand{\bra}[1]{\left\langle #1 \right|}
\newcommand{\ket}[1]{\left| #1 \right\rangle}

\title{
Towards the glueball spectrum
from unquenched lattice QCD.
}

\author[a,d]{E. Gregory}
\author[b]{A. Irving}

\author[c]{B. Lucini}
\author[d]{C. McNeile}
\author[e]{A. Rago}
\author[b]{C. Richards}

\author[f,g]{E. Rinaldi}

\affiliation[a]{
Department of Physics, University of Cyprus, P.O. Box
  20357
1678 Nicosia, Cyprus
}

\affiliation[b]{
Theoretical Physics Division, Dept. of Mathematical Sciences,
University of Liverpool, Liverpool L69 7ZL, UK
}

\affiliation[c]{
Department of Physics, College of Science, Swansea University,
  Singleton Park, Swansea SA2 8PP, UK
}

\affiliation[d]{
Bergische Universit\"at Wuppertal, Gaussstr.\,20, D-42119 Wuppertal,
Germany
}

\affiliation[e]{
School of Computing \& Mathematics, University of Plymouth,
Plymouth, PL4 8AA, UK 
}

\affiliation[f]{
  SUPA, School of Physics and Astronomy, University of Edinburgh,
  Edinburgh EH9 3JZ, UK
}

\affiliation[g]{
  Kobayashi-Maskawa Institute for the Origin of Particles and the Universe (KMI), Nagoya University, Nagoya 464-8602, Japan
}

\emailAdd{gregory@uni-wuppertal.de, aci@liverpool.ac.uk,
 b.lucini@swansea.ac.uk, mcneile@uni-wuppertal.de,
 antonio.rago@plymouth.ac.uk, cmr@liverpool.ac.uk,
 e.rinaldi@sms.ed.ac.uk
}

\abstract{
We use a variational technique to study heavy glueballs
on gauge configurations generated with 2+1 flavours
of ASQTAD improved staggered fermions.
The variational technique includes glueball scattering
states.
The measurements were made using $2150$ configurations at $0.092$ fm
with a pion mass of $360$ MeV.  We report masses for 10 glueball
states. We discuss the prospects for unquenched lattice QCD
calculations of the oddballs. 
}

\begin{document}

\maketitle

\section{Introduction and motivation}

Nothing is more symbolic of the difficulty of solving
QCD, than the fact that, while glueballs are central to
the understanding of non-perturbative QCD, there is
currently no definite experimental evidence for
their existence.
After much work the glueball spectrum~\cite{Bali:1993fb}
in quenched QCD
was mapped out by 
Morningstar and 
Peardon and collaborators~\cite{Morningstar:1999rf,Chen:2005mgo}. 
Results for higher spin glueballs have been reported by
Meyer and Teper~\cite{Meyer:2004jc,Meyer:2003hy} (see 
also~\cite{Liu:2001wqa}).

There has been much less work done on studying
the effect of sea quarks on the glueball 
masses~\cite{Bali:2000vr,Hart:2001fp,Hart:2006ps,Richards:2010ck}.
Glueball calculations suffer from a severe problem
with the signal to noise ratio that requires high
statistics. Also, some of the more sophisticated algorithms 
used in quenched
QCD calculations
that improve the signal to noise error, 
do not work for 
unquenched calculations~\cite{Meyer:2003hy,DellaMorte:2010yp}.

The properties of glueballs can be elucidated
by studying the experimental 
decay and production of flavour 
singlet mesons. Some of these analyses find that
the glueball degrees of freedom have similar masses
to the quenched 
glueballs~\cite{Cheng:2006hu}, while others find very different
masses~\cite{Cheng:2008ss,Mennessier:2008kk}. 
For example, one analysis~\cite{Cheng:2008ss} of the 
decay properties of the $0^{-+}$ states
suggested that large unquenching effects moved
the quenched $0^{-+}$ glueball from 2.6 GeV to 
1.4(1) GeV, close
to the experimental mass of the $\eta(1405)$ meson.

Some have argued (e.g.~\cite{Close:2001ga}) that there is an additional state
over the two expected $\overline{q} q$ mesons
in the $0^{++}$ flavour singlet mesons:
$f_0(1370)$, $f_0(1500)$, and  $f_0(1710)$.
The mass of the $0^{++}$ glueball 
inferred from quenched QCD~\cite{Morningstar:1999rf}
is 1730(50)(80) MeV. This could be important in 
understanding the above mesons if one assumes that 
unquenching effects in the $0^{++}$
glueball are small. 
Other groups, e.g.~\cite{Mennessier:2008kk},
have argued that
unquenching of the $0^{++}$ may be large and that the
mass in QCD is close to the $\sigma$.
There is also some controversy as to whether the 
the $f_0(1370)$ is a real meson state~\cite{Ochs:2010cv}, although it
is listed in the Particle Data Group (PDG) summary tables.
Lattice calculations will eventually have to deal with 
the decays of the $f_0$ mesons and possible  coupling
of the $f_0(980)$ and $f_0(600)$.

The mixing of the glueball degrees of freedom with
flavour singlet mesons has meant that many lattice
groups~\cite{Aoki:2007rd,Jansen:2009hr,Feng:2010es,Lang:2011mn,Prelovsek:2011im}
have been studying the $\rho$ meson and P-wave
states such as the $a_1(1260)$, $b_1(1235)$ meson using a variety
of techniques so as to understand how to deal with resonances
in lattice QCD calculations.

So, in summary, there are no hadrons where
glueball degrees of freedom have been confirmed.
Looking to the future, there are ongoing experiments that are searching
for glueball degrees of freedom, such as the
BES III experiment~\cite{Asner:2008nq}. 
The BES III experiment has 
preliminary~\cite{Olsen:2012xn} results
for two new states with masses close to the 
$0^{+-}$
and $1^{+-}$ quenched glueball masses.
In 2018, the PANDA 
experiment~\cite{Lutz:2009ff} will 
search for heavy glueballs with masses under
5.4 GeV.
In particular they will look for oddball 
glueballs with exotic $J^{PC}$ quantum numbers
($0^{+-}$, $2^{+-}$,  $1^{-+}$, $0^{--}$ and $3^{-+}$) 
which are not allowed in quark models for quark-antiquark
mesons. In a quenched QCD calculation, 
Morningstar and Peardon~\cite{Morningstar:1999rf}
found two glueballs with exotic $J^{PC}$ = $2^{+-}$ and
$0^{+-}$ with masses 4140(50)(200) MeV and 4740(70)(230) MeV
respectively. As we review later, other quenched glueball
studies have not seen these states~\cite{Meyer:2004gx}.

One motivation for studying glueballs heavier than 3 GeV is
that there could be reduced mixing with flavour singlet
quark states. There have been speculations that there
are no further, or a reduced number of, light meson states 
above 3.1 GeV~\cite{Brisudova:1999ut,Page:2001gs},
based on string breaking of the heavy quark potential.
Swanson~\cite{Swanson:2005rc} critiques the use of screened potentials in hadron
spectroscopy.
The heaviest light meson in the PDG summary 
table~\cite{Nakamura:2010zzi}
is the
$f_6(2510)$ with a mass of 2.469 GeV. It is 
possible that there may be no experiments capable of producing
light mesons beyond 2.5 GeV. It could be that heavier mesons
made from light quarks have large widths, so it becomes
difficult to extract the masses from experiment.
Even though the signal to noise ratio is worse for heavier
glueballs than for light glueballs, it would simplify
lattice QCD calculations if heavier glueballs do not
mix with quark degrees of freedom, because it would be
simpler to identify a pure glueball state.
For glueball masses above 3 GeV, there is the possibility
that the glueballs may mix with charmonium states.
However, Page~\cite{Page:2001gs} suggests that the 
mixing between charmonium states
and glueballs will be small, because the charm quark
is heavy and hence it is difficult to excite a charm loop.

The hadron spectrum collaboration~\cite{Dudek:2011tt}
has computed
the spectrum of excited mesons and baryons, although
typically at a single lattice spacing and heavy
quark masses. They don't report any meson masses above
2.8 GeV. Studying such masses is already an impressive achievement,
and it is not clear whether they can investigate the
meson spectrum at even higher masses.

Hagedorn conjectured that the density of 
light hadrons goes like $e^{m/T}$ where
$m$ is the mass of the light hadrons
and $T$ is a 
constant~\cite{Hagedorn:1965st,Hagedorn:1968zz}.
Cohen and Krejcirik have recently
reviewed~\cite{Cohen:2011cr}
the evidence 
that the number of light hadrons
agrees with the 
Hagedorn conjecture.

There may be indirect ways of determining whether
there exist light mesons in the regime 3 to 6 Gev,
using information from thermodynamic studies.
For example, the hadron resonance gas model (HRGM) 
is used
in the phenomenology of heavy ion experiments and 
has successfully reproduced some results from
lattice QCD calculations at non-zero temperature.
The HRGM depends on the number of mesons and baryon
states. Typically the meson and baryon states 
listed in the PDG are used,
however Chatterjee et al.~\cite{Chatterjee:2009km}
have studied a different density of states, with
mesons with masses higher than those listed in the PDG.
Majumder and Muller use the results 
from lattice QCD~\cite{Borsanyi:2010cj,Borsanyi:2010bp}
calculations with the HRGM to claim the existence of 
new light resonances~\cite{Majumder:2010ik}.
Megias et al.~\cite{Megias:2012hk} study the effects 
of the additional hadrons predicted by the quark model
on the results of the HRGM.

The PANDA experiment is performing 
some Monte Carlo studies
of glueball production using glueball decay
widths of 10 MeV~\cite{Wiedner:2011mf}.
It will be important to check some of these
speculations about the density of light mesons
and the mixing of glueball and charmed hadrons.
                  
Oddball glueballs will not mix with 
light mesons with non-exotic $J^{PC}$ quantum numbers.
Even if the heavier glueball states do not mix
with quark degrees of freedom, there is the possibility
that the glueballs decay to other glueballs. The calculation
reported here explicitly includes glueball scattering states to check
for this.

In~\cite{Richards:2010ck}, the results for $0^{++}$, $0^{-+}$, and $2^{++}$
were presented on the gauge configurations also used in this
study\footnote{In particular, this work uses the set of configurations
  referred to as the {\em fine ensemble} in~\cite{Richards:2010ck}.}. 
A powerful 
variational code was developed in~\cite{Lucini:2010nv} that
included two glueball scattering states
and was used to study glueballs
in $SU(N)$ quenched QCD.
In this paper, we use this code to extend the study
in~\cite{Richards:2010ck}
to look at glueballs with new $J^{PC}$.
We will mostly study the glueballs with $J^{PC}$ quantum
numbers not studied in the earlier work~\cite{Richards:2010ck}.

There are a number of reviews on
the theory of glueballs~\cite{Klempt:2007cp,Mathieu:2008me},
lattice QCD calculations of glueballs~\cite{McNeile:2008sr},
and experimental
searches~\cite{Crede:2008vw} for glueball degrees of freedom.

\section{Details of the lattice QCD calculation}

\begin{table*}[tb]
\centering
\begin{tabular}{|c|c|c|c|c|c|c|} \hline
$\beta$ &  $L^3 \times T$  &  $a m_l$ & $a m_s$   &
$N_{cfg}$ & $N_{traj}$ & $r_0/a$ \\ 
\hline
7.095 &  $32^3 \times 64$ & 0.00775 & 0.031  & 2150 &
12900 & 5.059(10)
\\ \hline
\end{tabular}
\caption{Summary of the parameters 
of the ensemble 
used in this calculation. We use the 
same convention for the quark masses as used by the MILC 
collaboration. The configurations used are taken every $6$
trajectories. 
}
\label{tb:ensemble}
\end{table*}

The unquenched lattice QCD calculations used the ASQTAD 
improved staggered fermion action~\cite{Orginos:1998ue,Orginos:1999cr}
and the one link Symanzik
improved gauge 
action~\cite{Alford:1995hw}. These are the same actions used by the
MILC collaboration~\cite{Bazavov:2009bb}. 
The configurations were described in one paper 
on glueball masses~\cite{Richards:2010ck}
and one on the 
masses of the $\eta$ and $\eta^\prime$ mesons~\cite{Gregory:2011sg}.
The parameters
of the calculation are shown in Tab.~\ref{tb:ensemble}.
The lattice spacing $a$ was determined from the $r_0/a$ parameter
calculated from the heavy quark potential and 
the value of $r_0$ ($0.4661$ fm) determined
by the HPQCD collaboration~\cite{Davies:2009tsa}
on the configurations generated by the 
MILC collaboration~\cite{Bazavov:2009bb}.

Glueball masses were extracted with a variational calculation. The
variational technique used in this work was described in~\cite{Lucini:2010nv}.
Here we provide a brief summary. The basis states for the variational
procedure are made from operators that transform under an irreducible
representation of the cubic lattice group and have specified parity
and charge conjugation quantum numbers. The irreducible
representations of the cubic group, which is the discrete rotational
symmetry group of the lattice, are conventionally called $A_1$,
$A_2$, $E$, $T_1$ and $T_2$. Representations $A_1$ and $A_2$ have dimension 1, $E$
has dimension 2 and $T_1$ and $T_2$ have dimension 3. Below, we show
how the continuum spin can be obtained from mass eigenstates
transforming according to an irreducible representation of the cubic group.

\begin{figure}
  \centering
  \includegraphics[width=0.7\textwidth]{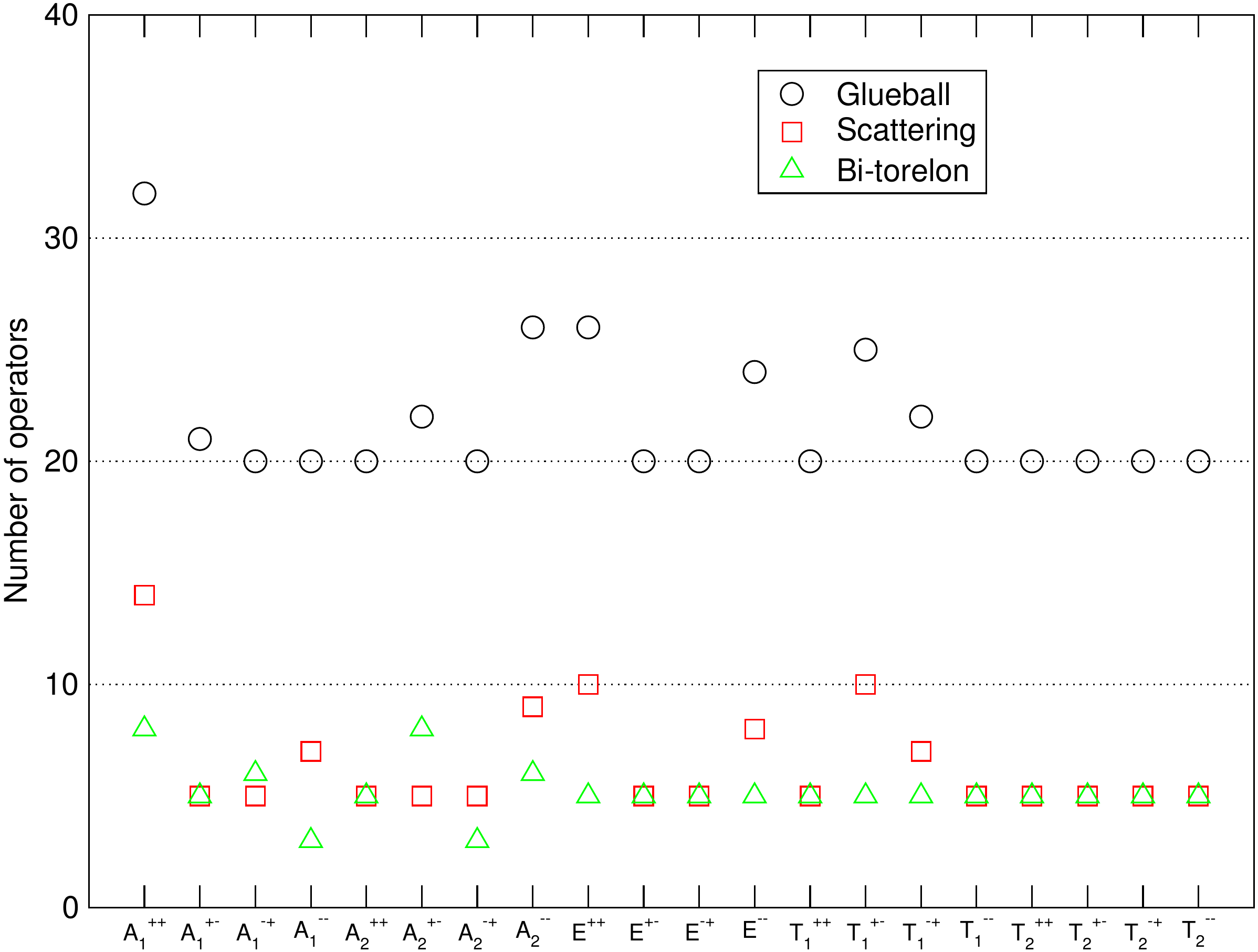} 
  \caption{Number of basis operators (before blocking and smearing) in
    each channel, for single glueballs, scattering states and
    bi-torelon states.}
  \label{fig:number_of_operators}
\end{figure}

On a given timeslice, a single glueball operator with well-defined
rotational quantum number is a prescribed linear combination of traced Wilson
loops of a given shape. Eigenstates of parity are obtained by considering
the starting linear combination and the reflected one, while
eigenstates of charge conjugations are given by the real part ($C =
1$) and the imaginary part ($C = -1$) of that combination. In our
calculations, we considered operators containing shapes from length 4 to
length 10. In addition to single glueball operators, the basis states 
include two-glueball scattering states (which involve products of two
basic shapes) and bi-torelon operators (products of two loops winding in
a compact spatial direction; note that these operators are not
necessarily straight lines). In order to keep 
the number of operators under control, we limited the number of 
single glueball operators to at most 30. The shapes we included in our calculations
are those that contribute in as many channels as possible and, barring
shapes that for symmetry reasons do not contribute to given
channels, the same basis shapes have been used for all the quantum
number assignments. Other
constraints imposed on the basis operators were that the variational
basis must contain shapes of all lengths and that a given shape can
contribute at most once in a given channel. This allowed us to include
a broad range of shapes and lengths in our basis. A Mathematica program
systematically constructed the basis operators with the requested
constraints. A summary of the number of operators we have used for
each quantum number assignment is reported in Fig.~\ref{fig:number_of_operators}.
Additional operators were generated by an improved
blocking algorithm~\cite{Lucini:2004my}. We used 6 different blocking
and smearing levels, which implies that for each channel our basis is 6
times larger than the number of operators quoted in
Fig.~\ref{fig:number_of_operators}. 

Consider two basis operator $\Phi_\alpha$  and $\Phi_\beta$. In the
lattice QCD calculations the correlators between operators separated
by time $t$ are computed and define the correlation matrix 
\begin{equation}
  \label{eq:corr-matrix}
  \tilde{C}_{\alpha\beta}(t) \; = \; \sum_\tau \bra{0}
  \Phi_\alpha^\dag(t+\tau) \Phi_\beta(\tau) \ket{0}
  \quad .
\end{equation}

The masses and amplitudes of glueball states can be extracted from
the generalized eigenvalue problem on the measured correlation
matrix: the optimal operators (i.e. those that almost coincide with
pure states) are the eingenvectors of 
\begin{equation}
  \label{eq:diagonalization}
  \bar{C}(\bar{t})=\tilde{C}^{-1}(0) \tilde{C}(\bar{t}) \ .
\end{equation}
These interpolating operators are a linear sum of the basis vectors: 
\begin{equation}
  \label{eq:optimal-op}
  \tilde{\Phi}_{i} (t) \; = \; \sum_\alpha v^i_\alpha \Phi_\alpha(t)
  \quad .
\end{equation}

The mass of the state is extracted by $\cosh$ fits
of the correlation matrix in the optimal basis:
\begin{equation}
\label{eq:coshFIT}
\bar{C}_{ii}(t) = |c_i|^2 \cosh \left( m_i t - N_T/2 \right)
\quad ,
\end{equation}
where $N_T$ is the length of the lattice in the time
direction and the $\cosh$ functional form is a consequence of the
usual exponential decay in a lattice with periodic boundary condition
in the time direction. 

\begin{figure}
  \centering
  \includegraphics[width=0.7\textwidth]{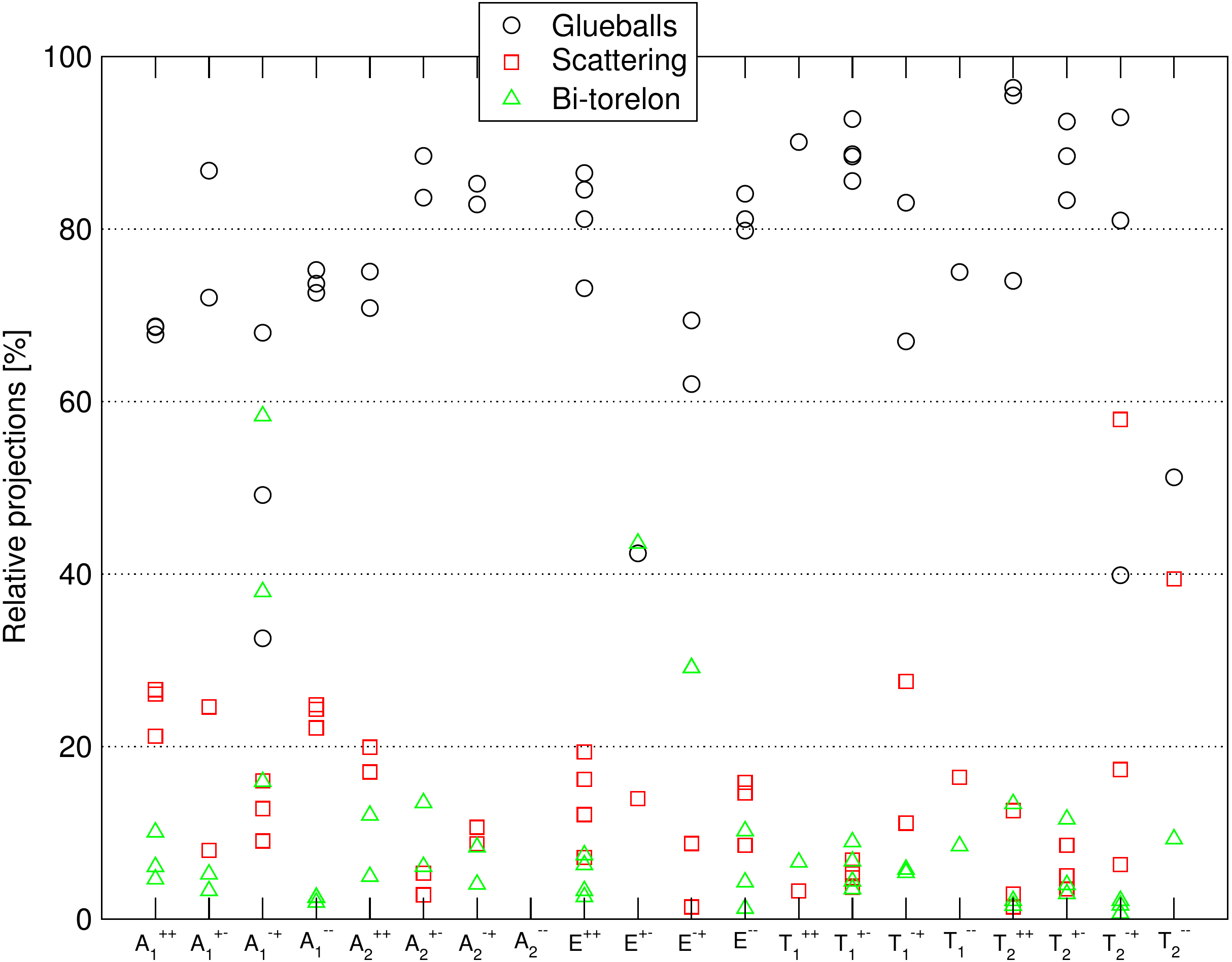} 
  \caption{Summary of the projections relative to the subspace spanned
  by different types of basis operators for the lightest states in the
spectrum.}
  \label{fig:relative-projections}
\end{figure}

In general, glueball correlators are very noisy and this limits the
usefulness of numerical correlators to short time separations. However,
although Eq.~(\ref{eq:coshFIT}) is only valid at large $t$, if the
overlap with an Hamiltonian state is almost perfect, it is possible to
extract a reliable value for the mass at short time separation, since
the decay is largely dominated by a single state. For this to be true, a
careful construction of the variational basis is paramount. Whether
an optimal state $\tilde{\Phi}_{i}$ is a good approximation of the Hamiltonian
eigenstates can be checked by looking at the value of the overlap $|c_i|^2$:  the
closer this number is to one from below (with one being the unitarity
limit), the better is the variational
calculation. In addition, it is important to estimate the
contributions to the mass coming from scattering and torelon states.
For the eigenstate corresponding to a true spectral mass, these
contaminations must be absent. The contribution of scattering
and torelon states for a given optimal state can be resolved by
looking at the relative length of its projection onto the space spanned
respectively by the scattering and bi-torelon basis operators. A
summary of the resulting relative projections is shown in
Fig.~\ref{fig:relative-projections}. For the 
lightest states extracted in each symmetry channel we show the
relative projection onto the different types of basis operators
(single glueballs, scattering and bi-torelon operators).

We do not include any fermionic scattering states in 
this calculation.  Fu has recently 
reported~\cite{Fu:2012gf} preliminary results 
for the decay width of the $0^{++}$ $\sigma$ into $\pi\pi$.
New techniques for calculating the masses of scalar mesons
using lattice QCD have been 
developed~\cite{Bernard:2010fp,Doring:2012eu}.

The results come from bin sizes of 50 configurations.
We have checked for autocorrelation effects by choosing different
binning blocks down to 5 configs. We found no evidence for
correlations.

\begin{figure}
  \centering
  \includegraphics[width=0.7\textwidth]{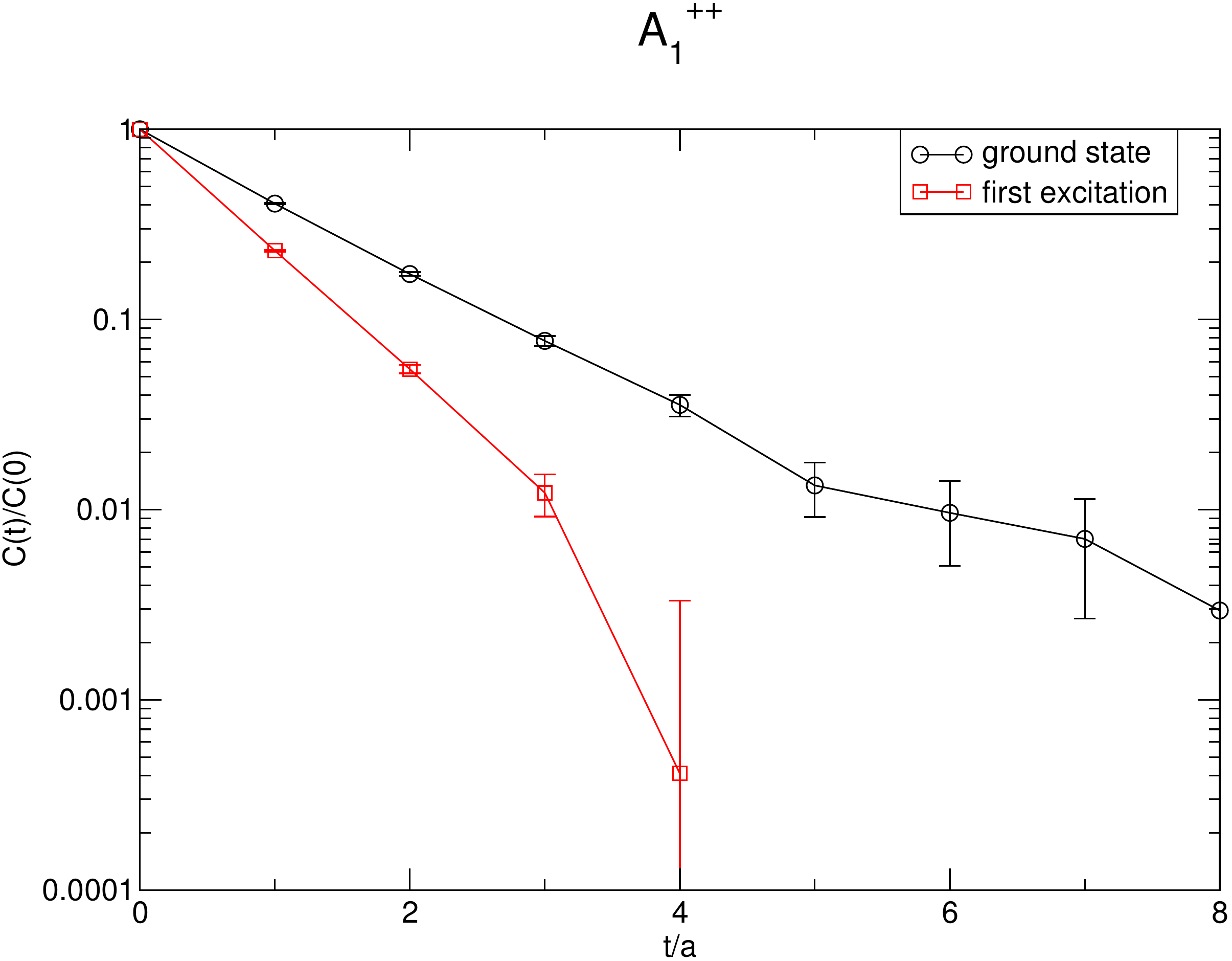} 
  \caption{Normalized correlator in log scale for the two lightest states in the
spectrum of the $0^{++}$ channel.}
  \label{fig:0RPpCp-corr}
\end{figure}

\begin{figure}
  \centering
  \includegraphics[width=0.7\textwidth]{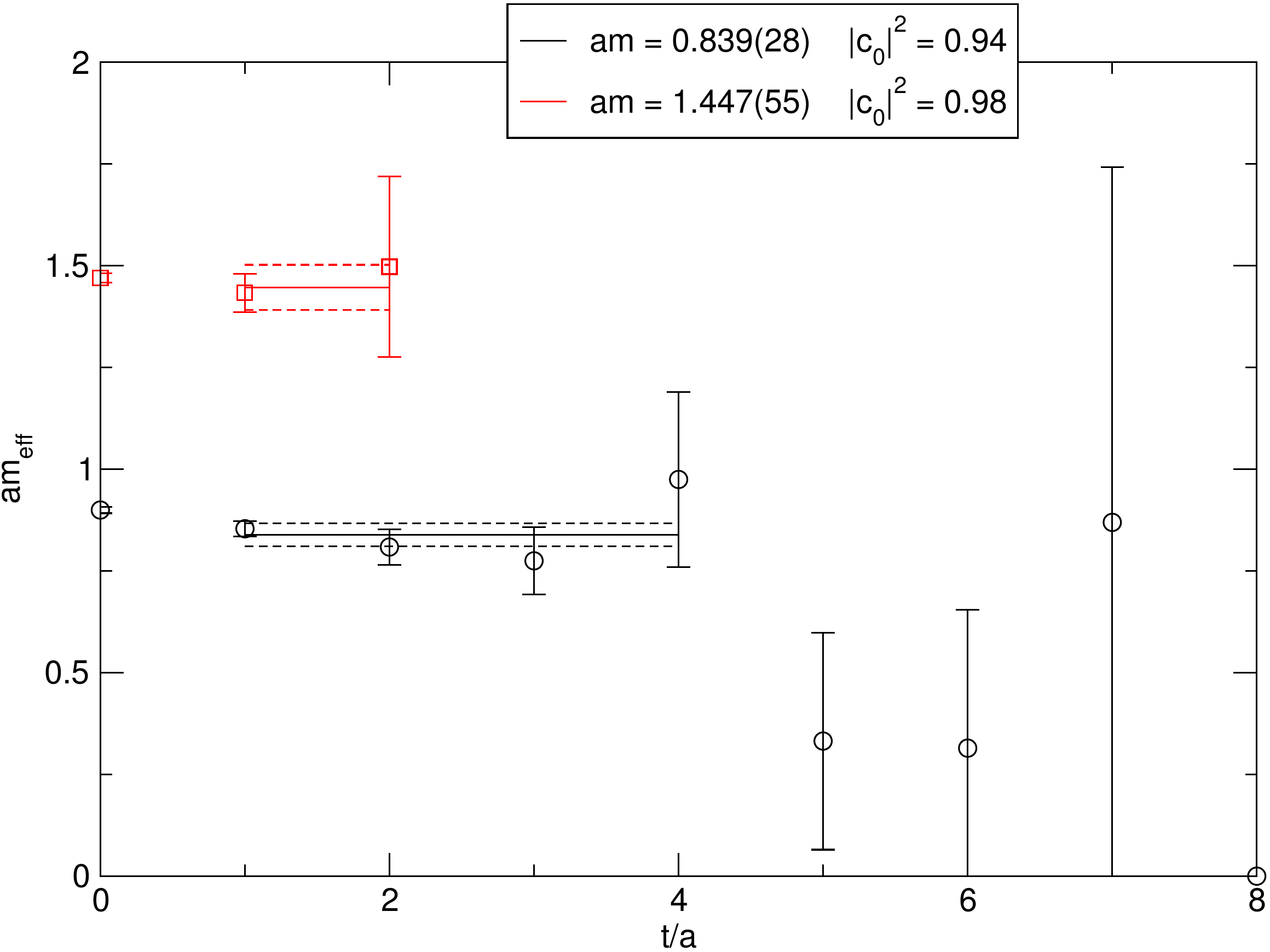} 
  \caption{Effective mass plateaux for ground state (circles) and the first
    excitation (squares) in the $0^{++}$
    channel. The fitted values are shown with solid
    lines (the dashed lines represent one standard deviation).}
  \label{fig:0RPpCp-mass}
\end{figure}

\begin{figure}
  \centering
  \includegraphics[width=0.7\textwidth]{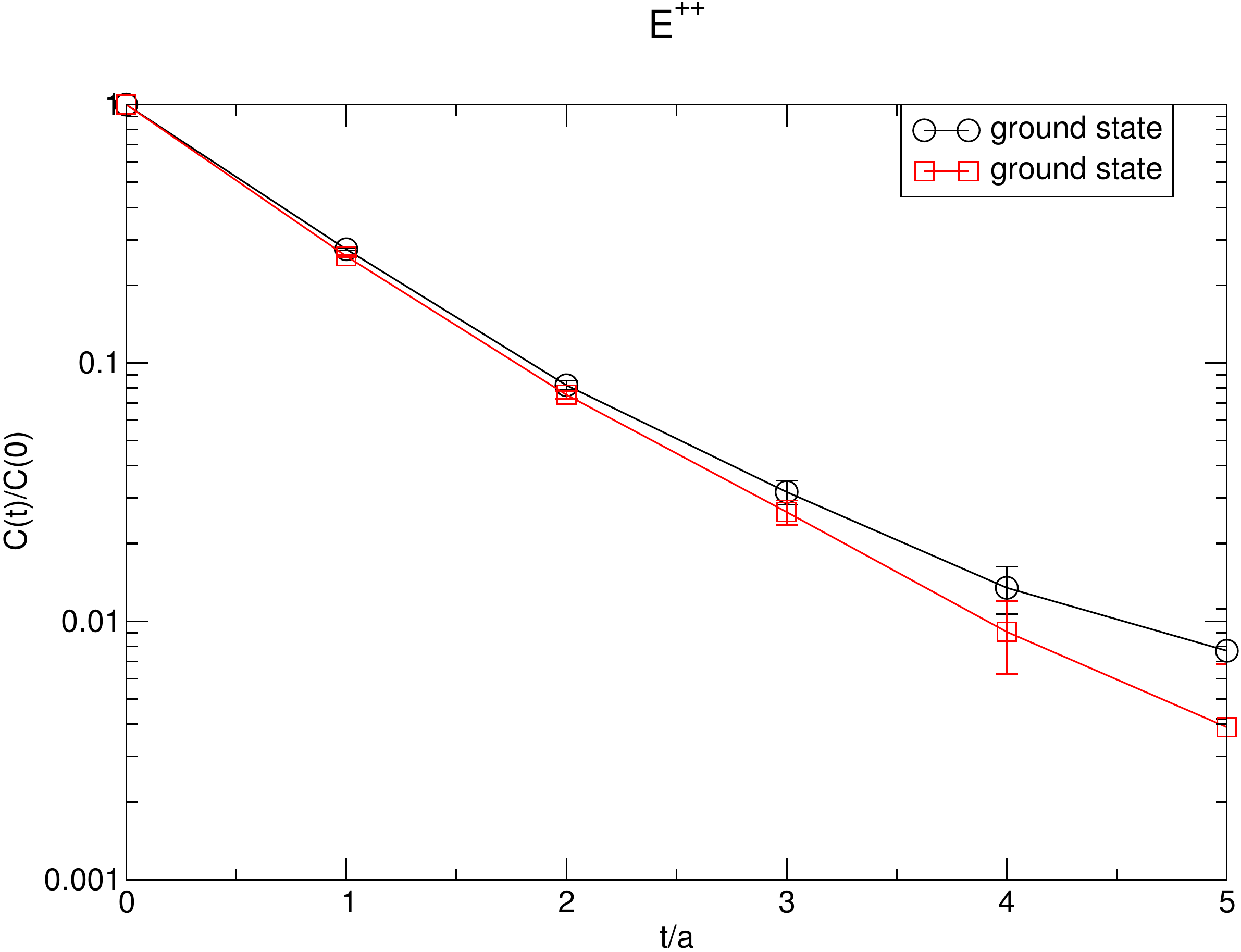} 
  \caption{Normalized correlator in log scale for the lightest state in the
spectrum of the $E^{++}$ channel.}
  \label{fig:ERPpCp-corr}
\end{figure}

\section{Results}
We reduced the variational correlators Eq.~\eqref{eq:corr-matrix} truncating the diagonal matrix Eq.~\eqref{eq:diagonalization} to the 5 highest eigenvalues (corresponding to the 5 lowest masses).
The diagonal correlators are fitted 
to the fit model in Eq.~\eqref{eq:coshFIT}.
The functional form is a good approximation if the overlap
$|c_i|^2 \sim O(1) $ 
and the $\chi^2/{\rm dof} \sim O(1)$.
All the states we were able to measure have 
$0.65 < |c_i|^2  < 1.1$,
with order 0.1 errors and $0 < \chi^2/{\rm dof} < 1.25$.

The correlator for the lightest glueball (the $A_1^{++}$ channel,
which in the continuum encodes the scalar $0^{++}$
glueball) and its first excitation is shown in
Fig.~\ref{fig:0RPpCp-corr} on a
logarithmic scale. The effective mass plateaux of
these two states with the corresponding estimate of the fitted masses from the fit in
Eq.~(\ref{eq:coshFIT}) are reported in Fig.~\ref{fig:0RPpCp-mass}.
A similar plot for the $E^{++}$ channel corresponding to the continuum
tensor glueball is shown in Fig.~\ref{fig:ERPpCp-corr}. For the
$A_1^{++}$, we have investigated alternative variational calculations
that include only some of the subsets of our operators (only single
glueball operators or only scattering and torellon operators). We note that
in both cases the mass of the ground state is the same within
errors. This phenomenon is different from what happens in pure gauge,
where for sets excluding single glueball operators the ground state
mass is roughly twice the mass of the ground state one finds when
single glueball are included in the variational basis~\cite{Lucini:2010nv}. This might be
due to the explicit breaking of centre symmetry in the presence of
dynamical fermions. Although the coupling of glueball states and
single torellon states was found to be negligible
in~\cite{Hart:2001fp}, it is possible that it becomes more relevant
closer to the chiral limit. To investigate this coupling systematically
would require variational calculations with a basis including single
torellon operators at various quark masses close to the chiral
limit. A similar calculation is beyond the scope of this paper.

\begin{figure}
\centering
\includegraphics[width=0.75\textwidth]{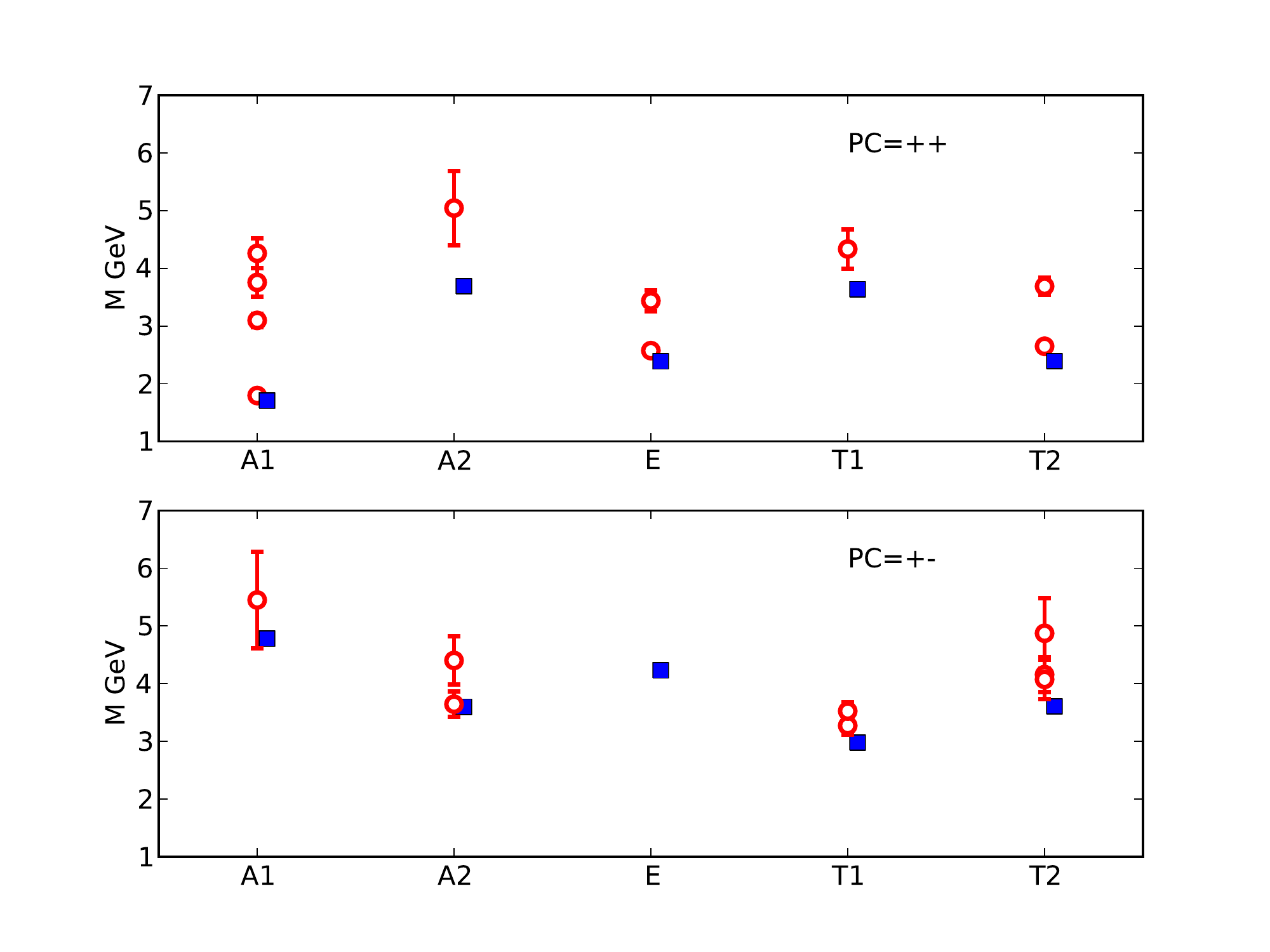}
~\\
\vspace{-0.5cm}
\includegraphics[width=0.75\textwidth]{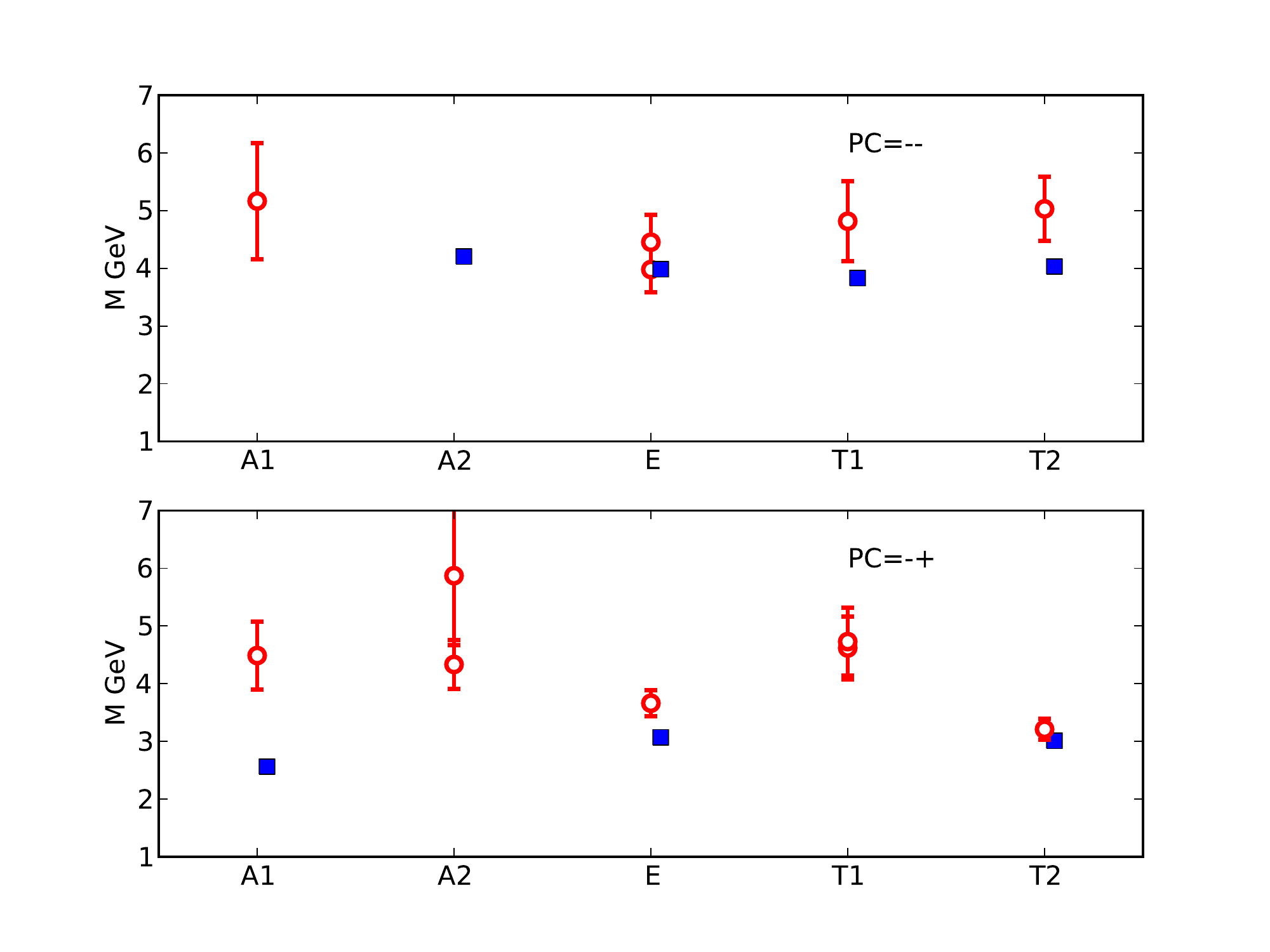} 
\caption{Glueball masses for the different lattice
  representations. Open circles refer to results obtained in this
  study, filled squares are quenched results from~\cite{Chen:2005mgo}.}
\label{fig:GlueMasses}
\end{figure}

In Fig.~\ref{fig:GlueMasses} we plot the glueball masses for each
lattice representation. A summary of the spectrum is reported in
Tab.~\ref{tab:spectrum-fine}. We define glueball states 
as those with an overlap of at least 65\% onto the single 
glueball operators.\\
We also include the results from 
Chen et al.~\cite{Chen:2005mgo} from a quenched lattice 
QCD calculation in the continuum limit. Fig.~\ref{fig:GlueMasses}
does not show any significant unquenching effects, although
the continuum limit of the unquenched results should be taken before a
definitive statement can be made.

\section{Classifying the glueballs into continuum angular momentum states}

Each of the states comes from an operator that is a linear combination of
ones in the variational basis. 
The projection on the 3 subsets can be used
to identify single glueball states from scattering or bi-torelon states. 
Only the $A_2^{--}$, $T_2^{-+}$ and $T_2^{--}$ channels had O(40)\% overlap
with the scattering subset of operators.
Also only the $A_1^{-+}$ and $E^{+-}$ channels
had contaminations of  O(40)\% overlap with bi-torelons.
We find that the channels
$A_1^{++}$, $A_1^{+-}$, $A_1^{--}$ and $T_1^{-+}$
had contaminations with scattering states
of between 20\% and 40\%.
The $A_1^{-+}$ and $E^{-+}$ channels
had bi-torelons contaminations between 20\% and 40\%.

\begin{table}[ht]
  \centering
  \begin{tabular}{|c|c|c|c|c|c|c|c|}
    \hline
    Ch. & $t_{\rm min}-t_{\rm max}$ & $am$ & $|c_0|^2$ & $\chi^2/{\rm dof}$ & ${\rm mix}_G$ & ${\rm mix}_S$ & ${\rm mix}_T$ \\
    \hline    \hline
    $A_1^{++}$ & 1-5 & 0.839(28) & 0.94(3) & 0.23 & 0.6874 & 0.2661 & 0.0466 \\
    $A_1^{++\star}$ & 1-5 & 1.447(55) & 0.98(5) & 0.48 & 0.6779 & 0.2611 & 0.0609 \\
    $A_1^{++\star\star}$ & 1-4 & 1.75(11) & 0.88(10) & 0.91 & 0.6865 & 0.2124 & 0.1010 \\
    $A_1^{++\star\star\star}$ & 1-5 & 1.99(12) & 1.16(14) & 0.91 & 0.6727 & 0.2055 & 0.1217 \\
    \hline
    $A_1^{+-}$ & 1-5 & 2.55(39) & 0.99(38) & 0.31 & 0.7207 & 0.2461 & 0.0331 \\
    \hline
    $A_1^{-+}$ & 1-6 & 1.30(4) & 0.89(3) & 0.15 & 0.3257 & 0.0908 & 0.5835 \\
 $A_1^{-+\star}$ & 1-5 & 1.57(8) & 0.73(6) & 0.79 & 0.4919 & 0.1284 &
    0.3797 \\
 $A_1^{-+\star\star}$ & 1-5 & 2.10(28) & 0.76(20) & 0.97 & 0.6800 & 0.1605 & 0.1595 \\
    \hline
    $A_1^{--}$ & 1-4 & 2.41(47) & 0.81(39) & 0.64 & 0.7263 & 0.2486 & 0.0251 \\
    \hline
    $A_2^{++}$ & 1-4 & 1.94(13) & 0.99(13) & 0.30 & 0.7509 & 0.1995 & 0.0495 \\
    \hline
    $A_2^{++}$ & 1-4 & 2.36(30) & 1.08(33) & 0.06 & 0.7086  & 0.1708 & 0.1206 \\
    \hline
    $A_2^{+-}$ & 1-5 & 1.70(10) & 0.76(8) & 0.48 & 0.8366 & 0.0284 &
    0.1350 \\
    $A_2^{+-\star}$ & 1-3 & 2.06(10) & 0.79(15) & 0.41 & 0.8850 & 0.0535 & 0.061 \\
    \hline
    $A_2^{-+}$ & 1-4 & 2.03(20) & 0.66(13) & 0.56
 & 0.8528  & 0.1067 & 0.0405 \\
    $A_2^{-+\star}$ & 1-4 & 2.74(56) & 0.97(60) & 1.03 & 0.8528 & 0.1067 & 0.0405 \\
    \hline
    $A_2^{--}$ & 1-5 & 2.32(25) & 0.98(25) & 1.10 & 0.5526 & 0.3905 & 0.0569 \\
    \hline
    $E^{++}$ & 1-5 & 1.202(32) & 0.86(3) & 0.97 & 0.8117 & 0.1622 & 0.0261 \\
    $E^{++\star}$ & 1-5 & 1.606(87) & 0.89(8) & 0.44 & 0.8457 & 0.1213 & 0.0330 \\
    \hline
    $E^{+-}$ & 1-4 & 2.42(40) & 0.98(39) & 0.01 & 0.4243 & 0.1398 & 0.4359 \\
    \hline
    $E^{-+}$ & 1-5 & 1.42(6) & 0.76(5) & 0.76 & 0.4243  & 0.1398 &
    0.4359 \\
    $E^{-+\star}$ & 1-3 & 1.71(10) & 0.81(8) & 0.57 & 0.6943 & 0.0143 & 0.2914 \\
    \hline
 $E^{--}$ & 1-5 & 1.86(19) & 0.71(13) & 1.1 & 0.7984 & 0.1586 & 0.0430 \\
$E^{--\star}$ & 1-5 & 2.08(22) & 0.83(18) & 0.92 & 0.8411 & 0.1464 & 0.0126 \\
    \hline
    $T_1^{++}$ & 1-4 & 2.03(16) & 0.92(15) & 0.90 & 0.9011 & 0.0327 & 0.0662 \\
    \hline
$T_1^{+-}$ & 1-5 & 1.530(71) & 0.86(6) & 0.45 & 0.8870 & 0.0689 & 0.0441 \\
$T_1^{+-\star}$ & 1-4 & 1.65(7) & 0.87(6) & 0.26 & 0.9277 & 0.0371 & 0.0352 \\
    \hline
    $T_1^{-+}$ & 1-5 & 2.16(26) & 0.89(23) & 0.24 & 0.6700 & 0.2759 & 0.0541 \\
$T_1^{-+\star}$ & 1-5 & 2.21(28) & 0.85(24) & 1.15 & 0.8308 & 0.1114 & 0.0578 \\
    \hline
    $T_1^{--}$ & 1-5 & 2.25(32) & 0.96(31) & 0.33 & 0.7504 & 0.1645 & 0.0851 \\
    \hline
$T_2^{++}$ & 1-4 & 1.238(43) & 0.92(4) & 0.63 & 0.9640 & 0.0144 & 0.0216 \\
$T_2^{++\star}$ & 1-4 & 1.723(70) & 0.92(6) & 0.85 & 0.9551 & 0.0291 & 0.0159 \\
    \hline
$T_2^{+-}$ & 1-4 & 1.90(16) & 0.83(13) & 0.99 & 0.8336 & 0.0504 & 0.1160 \\
$T_2^{+-\star}$ & 1-4 & 2.28(29) & 0.94(26) & 0.92 & 0.9249 & 0.0350 & 0.0401 \\
    \hline
$T_2^{-+}$ & 1-4 & 1.50(8) & 0.76(6) & 0.93 & 0.8101 & 0.1736 & 0.0163 \\
$T_2^{-+\star}$ & 1-5 & 1.69(12) & 0.90(10) & 0.28 & 0.3989 & 0.5795 &
    0.0217 \\
    \hline
$T_2^{--}$ & 1-4 & 2.35(26) & 1.07(26) & 0.91 & 0.5124 & 0.3946 & 0.0930 \\
    \hline
    \hline
  \end{tabular}
  \caption{Spectrum on the ensemble of Tab.~\ref{tb:ensemble}. For
    each state we were able to extract a signal, we show the
    parameters of the fit with Eq.~\eqref{eq:coshFIT} and the relative
    projection of the mass eigenstate on the different subsets of
    operators (${\rm mix}_G$, ${\rm mix}_S$, ${\rm mix}_T$). Good
    overlaps $|c_0|^2 \sim {\cal O}(1)$ and $\chi^2/{\rm dof}$ are
    shown.}
  \label{tab:spectrum-fine}
\end{table}

\begin{table}[ht]
\begin{center}
  \begin{tabular}[h]{c|ccccc}
    $J$ & $A_1$ & $A_2$ & $E$ & $T_1$ & $T_2$ \\
    \hline
    0 & 1 & 0 & 0 & 0 & 0 \\
    1 & 0 & 0 & 0 & 1 & 0 \\
    2 & 0 & 0 & 1 & 0 & 1 \\
    3 & 0 & 1 & 0 & 1 & 1 \\
    4 & 1 & 0 & 1 & 1 & 1 \\
  \end{tabular}
\end{center}
\caption{Subduced representations $J \downarrow \mathcal{G}_O$ of the octahedral group up to $J=4$.
   This table illustrates the spin content of the irreducible
   representations of $\mathcal{G}_O$ in terms of the continuum $J$.}
\label{tab:subd-reps}
\end{table}

\begin{table}[ht]
\begin{center}
  \begin{tabular}[h]{|c|c|c|c|c|}
\hline
$J^{PC}$  & \multicolumn{4}{|c|}{Mass MeV} \\
\cline{2-5}
 & Unquenched & \multicolumn{3}{|c|}{Quenched } \\
\cline{3-5}
 & This work &  M\&P & Ky & Meyer\\  \hline
$0^{-+}$ &   & 2590(40)(130) & 2560(35)(120) & 2250(60)(100)\\ 
$2^{-+}$ & 3460(320) & 3100(30)(150) & 3040(40)(150) & 2780(50)(130) \\
$0^{-+}$ & 4490(590) & 3640(60)(180) &  & 3370(150)(150) \\ 
$2^{-+}$ & &               &  & 3480(140)(160) \\ 
$5^{-+}$ &           &               &  & 3942(160)(180) \\ \hline
$0^{--}$ (exotic) & 5166(1000)  &  &  &  \\
$1^{--}$ &   & 3850(50)(190) & 3830(40)(190) & 3240(330)(150) \\
$2^{--}$ & 4590(740) & 3930(40)(190) & 4010(45)(200) & 3660(130)(170) \\
$2^{--}$ &  &  &  & 3.740(200)(170) \\
$3^{--}$ &  & 4130(90)(200) & 4200(45)(200) & 4330(260)(200) \\ \hline
$1^{+-}$ & 3270(340) & 2940(30)(140) & 2980(30)(140) & 2670(65)(120) \\
$3^{+-}$ & 3850(350) & 3550(40)(170) & 3600(40)(170) & 3270(90)(150) \\
$3^{+-}$ &           &               &               & 3630(140)(160) \\
$2^{+-}$ (exotic)    &  & 4140(50)(200) & 4230(50)(200) & \\ 
$0^{+-}$ (exotic)    & 5450(830) & 4740(70)(230) & 4780(60)(230) & \\ 
$5^{+-}$  &          &           &               & 4110(170)(190) \\ \hline
$0^{++}$ & 1795(60)  & 1730(50)(80)   & 1710(50)(80)  & 1475(30)(65)\\
$2^{++}$ & 2620(50) & 2400(25)(120)  & 2390(30)(120) & 2150(30)(100) \\ 
$0^{++}$ & 3760(240) & 2670(180)(130) &  & 2755(30)(120) \\ 
$3^{++}$ &  & 3690(40)(180) & 3670(50)(180) & 3385(90)(150)\\ 
$0^{++}$ &  &  &  &  3370(100)(150) \\ 
$0^{++}$ &  &  &  &  3990(210)(180) \\ 
$2^{++}$ &  &  &  &  2880(100)(130) \\ 
$4^{++}$ &  &  &  &  3640(90)(160)  \\ 
$6^{++}$ &  &  &  &  4360(260)(200) \\ 
\hline
  \end{tabular}
\end{center}
\caption{Glueball masses with $J^{PC}$ assignments.
The column M\&P reports results from 
Morningstar and Peardon~\cite{Morningstar:1999rf}
from quenched QCD. The column labelled Ky is the data
from Chen et al.~\cite{Chen:2005mgo}.
Meyer's results are from~\cite{Meyer:2004gx}. 
}
\label{tab:JPC}
\end{table}

Identification of the spin of 
glueball states on the lattice is non-trivial. Continuum
spin representations break down into lattice representations at finite
lattic spacing.
In Tab.~\ref{tab:subd-reps} the continuum spin representations
are broken down into lattice representations.
\clearpage
Meyer and Teper~\cite{Meyer:2002mk,Meyer:2004gx} have
developed systematic techniques to classify glueball
masses into spins. 
Dudek et al.~\cite{Dudek:2007wv} have stressed the importance of 
correctly identifying
lattice representations with spin in the charmonium system.
Here, we use the simplest spin identification.
We identity the $J=0$ state with the $A_1$, and 
$J=1$ state with the $T_1$ representation, and
$J=2$ states with almost degenerate $T_2$ and $E$ representations.
For $J=2$ and $J=3$ states we take a weighted average 
of the masses in the component lattice representations.

Morningstar and Peardon~\cite{Morningstar:1999rf}
found glueball states with $0^{-+}$ (ground and excited), 
and $2^{-+}$ quantum numbers.
In the $PC=-+$ sector in Fig.~\ref{fig:GlueMasses} 
there are two almost degenerate levels in the $E$
and $T_2$ representations that could come from
a $J=2$ state. There is one level in the $A_1$ representation
which is relatively isolated and so is expected to couple to $0^{-+}$ states.
There are also degenerate levels in the $A_2$, and $T_1$
representations that could be $3^{-+}$. This would
be interesting, because $3^{-+}$ is an exotic quantum number,
however, because there is no signal in $T_2$ representation,
this state is not included in the summary tables.

In the $PC=--$ sector in Fig.~\ref{fig:GlueMasses}, the errors are
large because the masses are very heavy. 
Morningstar and Peardon~\cite{Morningstar:1999rf}
found glueball states with $1^{--}$,
$2^{--}$, and $3^{--}$. 
Here we see evidence for a $2^{--}$ state with near-degenerate
masses in the $E$ and $T_2$ channel.
The results in Fig.~\ref{fig:GlueMasses} are
not consistent with $3^{--}$, because there is
no signal in the $A_2$ channel.
There is also a hint of a signal for the
spin exotic $0^{--}$  state in the $A_1$ representation,
but unfortunately with large errors.

\begin{figure}
\begin{center}
\includegraphics[width=0.7\textwidth]{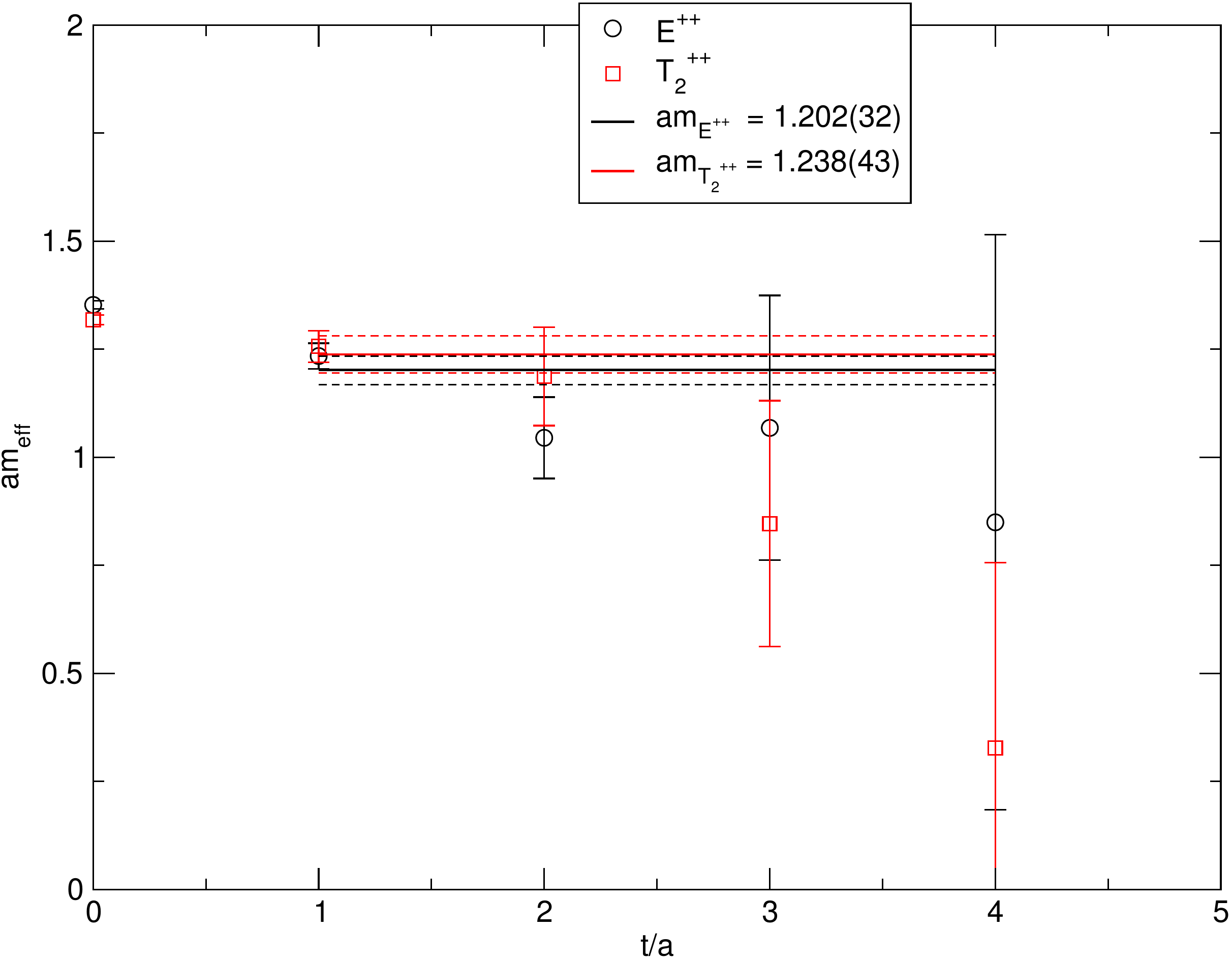}
\end{center}
\caption{The ground state effective mass in the $E^{++}$ channel is
  compared to the one in the $T_2^{++}$ channel. These channels
  correspond to the continuum spin $J=2$ representation and are
  supposed to be degenerate when the complete rotational symmetry is restored.
}
\label{fig:tensor-meff}
\end{figure}

For the operators with $PC=++$ in Fig.~\ref{fig:GlueMasses},
there is potentially a rich mixture of states. 
Morningstar and Peardon~\cite{Morningstar:1999rf}
found glueball states with $0^{++}$ (ground and excited),
and $2^{++}$.
There are two $0^{++}$
states in the $A_1$ channel. There are two almost degenerate states
in the $E$ and $T_2$ channel that would correspond to a $2^{++}$
state in the continuum. 
The agreement of the effective masses for these two lattice states,
together with the fitted masses using Eq.~(\ref{eq:coshFIT}) is shown
in Fig.~\ref{fig:tensor-meff}. The mass degeneracy between these two
channels is supposed to be exact in the continuum limit, when the full
three-dimensional rotational symmetry is dynamically restored.

There is potentially a $4^{++}$  
appearing in the 
$A_1$, $E$, $T_1$, and $T_2$ representations. There is also
a potential $3^{++}$ state in the $A_2$ representation,
but there is no hint of degenerate states in the $T_1$ and $T_2$
representations.

In quenched QCD, Morningstar and Peardon~\cite{Morningstar:1999rf}
found glueball states with $J^{PC}$ = $3^{+-}$, $1^{+-}$, $2^{+-}$,
and $0^{+-}$. 
The summary of the masses in Fig.~\ref{fig:GlueMasses}
doesn't show a signal in the $E^{+-}$, so our results
are inconsistent with an exotic $2^{+-}$ state. There
is evidence for states with $J^{PC}$ = $3^{+-}$, $1^{+-}$,
and $0^{+-}$.

\begin{figure}
\begin{center}
\includegraphics[scale=0.6]{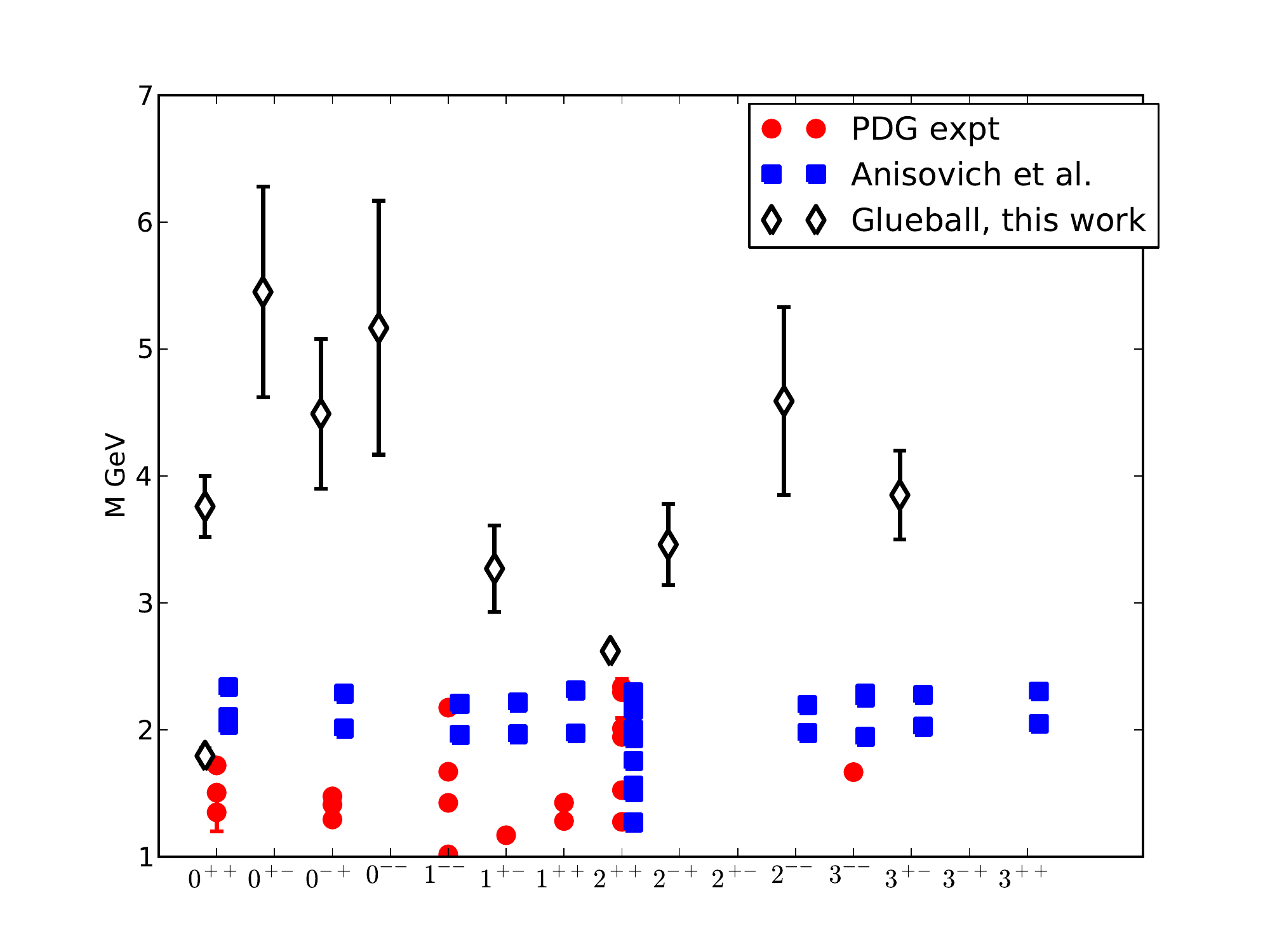}
\end{center}
\caption{
Summary of the glueball masses compared to experimental meson masses.
The experimental results are from the PDG~\cite{Nakamura:2010zzi}
and from Anisovich et al.~\cite{Anisovich:2011wx,Anisovich:2011sd,Anisovich:2011in}.
}
\label{fg:Finalsummary}
\end{figure}

In Tab.~\ref{tab:JPC} we summarise our assignments for
$J^{PC}$ quantum numbers to the glueball masses 
from operators, and compare with other results in the 
literature from quenched QCD calculations. 
Note that the
calculations by Meyer and Teper~\cite{Meyer:2002mk,Meyer:2004gx}
used the string tension to determine the lattice spacing,
however Morningstar and Peardon~\cite{Morningstar:1999rf}
and Chen et al.~\cite{Chen:2005mgo} used $r_0$. This may cause
a systematic difference in the results. Meyer~\cite{Meyer:2008gg} 
showed
that consistent results were obtained for the masses
of the tensor and scalar glueballs from the different
glueball calculations~\cite{Meyer:2002mk,Meyer:2004gx,Morningstar:1999rf,Chen:2005mgo},
if $r_0$ was always used to determine the 
lattice spacing.
In Fig.~\ref{fg:Finalsummary} we plot the summary
of the glueball masses.

To compare with experiment we use the masses
in the PDG. However we also include the additional 
masses reported by the Crystal Barrel collaboration,
reviewed by Bugg~\cite{Bugg:2004xu}. Note that
the Crystal Barrel experiment could only find new
states in a restricted mass range.
As explained in~\cite{Anisovich:2011in}, some of the 
results from 
Anisovich et al.~\cite{Anisovich:2011wx,Anisovich:2011sd,Anisovich:2011in}
are not included in the PDG summary tables,
because they are neither confirmed nor excluded
by other experiments. We plot the results for $I=0$ mesons
in~\cite{Anisovich:2011wx,Anisovich:2011sd},
but plot the $2^{++}$ states from~\cite{Anisovich:2011in}.
These additional meson states
from  Anisovich et
al.~\cite{Anisovich:2011wx,Anisovich:2011sd,Anisovich:2011in}
have been used to find some experimental 
glueball candidates~\cite{Bugg:2000zy} 
from the quenched calculations of Morningstar and Peardon.

Bali~\cite{Bali:2006xt} has plotted the 
quenched glueball spectrum with
the experimental masses of the charmonium system.

\section{Conclusions and future prospects}

The most conservative interpretation of our results
is that the masses in terms of lattice representations
are broadly consistent with results from quenched QCD.
We do not see any evidence for large unquenching effects,
however a definitive calculation requires a continuum extrapolation,
and the inclusion of fermionic operators. In
Tab.~\ref{fg:Finalsummary}
we tentatively assign $J^{PC}$ quantum numbers to 10
glueballs.

Of particular note in Tab.~\ref{tab:JPC} is that Meyer and
Teper~\cite{Meyer:2004gx} do not see the two spin exotic states identified 
by Morningstar and Peardon~\cite{Morningstar:1999rf}. 
In their summary of the glueball spectrum Morningstar
and Peardon note that their spin exotic glueball
 $2^{+-}$ could actually  be part of
$5^{+-}$, $7^{+-}$, or $11^{+-}$ glueball.
Mathieu~\cite{Mathieu:2008me} have also compared the results 
for glueball masses from Morningstar and Peardon with those
from Teper and Meyer.
Our result for the mass of the $0^{+-}$
are consistent with the result from
Morningstar and Peardon~\cite{Morningstar:1999rf},
although our errors are large for this heavy state. Meyer and
Teper\cite{Meyer:2004jc,Meyer:2004gx} used 
sophisticated measurement techniques to help assign
$J^{PC}$ quantum numbers to their results in terms of lattice
representations, but other groups reporting glueball masses
have not done this. Given the importance of the masses
of the oddball glueballs to the experimental program
of PANDA, this issue needs to be resolved.
Future lattice QCD calculations using heavy glueball degrees
of freedom should use improved techniques to assign
$J^{PC}$ quantum 
numbers~\cite{Liu:2001wqa,Meyer:2004jc,Meyer:2003hy}.

It would be advantageous to lattice QCD calculations if there were
no light mesons above 3.1 GeV, because this would reduce the
complicated mixing of glue and quark degrees of freedom. 
Aesthetically, it would be better to find an isolated glueball,
rather than have to determine the mixture of glueball and 
quark degrees of freedom in a state such as the $f_0(1790)$.
The 
determination of glueballs with exotic $J^{PC}$ may be easier
for lattice QCD calculations, although there is still the
problem of the heavy mass and the poor signal to noise ratio.

Unquenched glueball calculations require much higher statistics
than for the majority of flavour non-singlet
lattice QCD calculations. We can use the numbers 
in Tab.~\ref{tab:JPC} to estimate roughly 
the number of configurations required to achieve
a given accuracy. The most interesting states 
in Tab.~\ref{tab:JPC} are the ones with exotic
$J^{PC}$ = $0^{+-}$ and $2^{+-}$ quantum numbers. 
Using simple
$1/\sqrt{N_{\mbox{configs}}}$ scaling, we can estimate 
that we need 550,000 configurations
to get 50 MeV errors for the spin exotic $0^{+-}$ glueball.

The MILC collaboration has reported time estimates
to produce ensembles of gauge configurations
with 2+1+1 flavours of sea quarks with the
HISQ improved staggered action. With the light
quark a tenth of the strange quark mass, they
estimate the time to generate 1000 configurations
as 35, 128, and 352 Million core hours for
the lattice spacings of: 0.09, 0.06,
and 0.045 fm respectively\footnote{Talk presented by Paul Mackenzie at May 2011
  review
of the lqcd project.}.
These numbers suggest that
it is still too expensive to reduce the errors 
on the masses of the oddball glueballs. Lattice QCD
calculations that use anisotropic lattices may
be computationally cheaper.

The heavy glueballs may mix with charmonium states.
To study this type of mixing will probably require both
charm loops in the sea and the computation of disconnected
diagrams~\cite{McNeile:2004wu,Levkova:2010ft}
for the charmonium valence states. There are a number
of lattice QCD calculations, such as the one by the
MILC collaboration~\cite{Bazavov:2010ru}, 
that include the dynamics of the charm quarks
in the sea.

It is an exciting time for lattice QCD, with some
unquenched lattice QCD calculations done with
physical quark masses, and many unquenched calculations 
resulting in errors at the $O(1\%)$ level. The progress
on lattice QCD calculations of glueballs is
slower. However, the start of the PANDA experiment in 2018
provides an important deadline for lattice QCD calculations
of glueballs.

\acknowledgments{The calculations were performed on the Liverpool
cluster that is part of DiRAC funded by STFC.
The gauge configurations were generated on
the QCDOC~\cite{Boyle:2005gf}.
We used Chroma~\cite{Edwards:2004sx}
to make fermionic measurements. The work of B.L. is supported by the
Royal Society and by STFC. ER is supported by a SUPA prize studentship
and a JSPS short-term fellowship.}

\bibliographystyle{JHEP}
\bibliography{glueball}

\end{document}